\documentclass[iop]{emulateapj}
\pdfoutput=1

\usepackage{color}
\usepackage{times}
\usepackage{graphicx}
\usepackage{float}
\usepackage{ulem}
\usepackage{natbib}
\usepackage{microtype}
\usepackage{dcolumn}

\newcommand\checked[1]{#1}

\newcommand\unit[2]{\mbox{#1\,#2}}
\newcommand\mathunit[2]{#1\,#2}

\newcommand\timewindow{\unit{$[-5,\ +1)$}{s}}
\newcommand\seglen{\unit{6}{s}}

\newcommand\numberoff{324}

\newcommand\numGRBs{22}
\newcommand\numbertotaloff{6801}
\newcommand\mwuprob{53}

\newcommand\mediandistBNS{3.3}

\newcommand\mediandistBHNShigh{6.7}
\newcommand\statGRBLV{74.5}

\newcommand\mediandistNSBH{\mediandistBHNShigh}

\def\thercsid{\relax}
\def\rcsid#1{\def\next##1#1{\def\thercsid{##1}}\next}
\rcsid$Id: S5GRB.tex,v 1.191 2010/03/03 16:37:33 nvf Exp $

\renewcommand{\today}{\number\day\space\ifcase\month\or
  January\or February\or March\or April\or May\or June\or
  July\or August\or September\or October\or November\or December\fi
  \space\number\year}

\begin{document}

\title{Search for gravitational-wave inspiral signals associated with short Gamma-Ray Bursts during LIGO's fifth and Virgo's first science run}


\author{J.~Abadie\altaffilmark{29}, 
B.~P.~Abbott\altaffilmark{29}, 
R.~Abbott\altaffilmark{29}, 
T.~Accadia\altaffilmark{27}, 
F.~Acernese\altaffilmark{19ac}, 
R.~Adhikari\altaffilmark{29}, 
P.~Ajith\altaffilmark{29}, 
B.~Allen\altaffilmark{2,77}, 
G.~Allen\altaffilmark{52}, 
E.~Amador~Ceron\altaffilmark{77}
% ...
}

\affil{The LIGO Scientific Collaboration \& The Virgo Collaboration}

% ...
\altaffiltext{2}{Albert-Einstein-Institut, Max-Planck-Institut f\"ur Gravitationsphysik, D-30167 Hannover, Germany}
% ...
\altaffiltext{19}{INFN, sezione di Napoli $^a$; Universit\`a di Napoli 'Federico II'$^b$ Complesso Universitario di Monte S.Angelo, I-80126 Napoli; Universit\`a di Salerno, Fisciano, I-84084 Salerno$^c$, Italy}
% ...
\altaffiltext{27}{Laboratoire d'Annecy-le-Vieux de Physique des Particules (LAPP),  IN2P3/CNRS, Universit\'e de Savoie, F-74941 Annecy-le-Vieux, France}
% ...
\altaffiltext{29}{LIGO - California Institute of Technology, Pasadena, CA  91125, USA }
% ...
\altaffiltext{52}{Stanford University, Stanford, CA  94305, USA }
% ...
\altaffiltext{77}{University of Wisconsin--Milwaukee, Milwaukee, WI  53201, USA }
% ...

\begin{abstract}

Progenitor scenarios for short gamma-ray bursts (short GRBs) include coalescenses
of two neutron stars or a neutron star and black hole, which would necessarily
be accompanied by the emission of strong gravitational waves.  We present a search for
these known gravitational-wave signatures in temporal and directional
coincidence with \numGRBs{} GRBs that had sufficient gravitational-wave data
available in multiple instruments during LIGO's fifth science run, S5, and Virgo's first science run, VSR1\@.
We find no statistically significant gravitational-wave candidates
within a \timewindow{} window around the trigger time of any GRB\@.
Using the Wilcoxon--Mann--Whitney U test, we find no evidence for an
excess of weak gravitational-wave signals in our sample of GRBs.
We exclude neutron star--black hole progenitors to a median 90\% confidence exclusion
distance of \unit{\mediandistNSBH}{Mpc}.

\end{abstract}


\maketitle

\section{Introduction}

The past decade has seen dramatic progress in the understanding of
gamma-ray bursts (GRBs), intense flashes of $\gamma$-rays that are
observed to be isotropically distributed over the sky \citep[see e.g.
][and references therein]{Klebesadel:1973,meszaros:2006}.
The short-time variability of the bursts
indicates that the sources are very compact.  They are observed directly
by $\gamma$-ray and X-ray satellites in the Interplanetary Network (IPN) such as
HETE, Swift, Konus--Wind, INTEGRAL, and Fermi
\citep[see][and references therein]{2003AIPC..662....3R,Gehrels:2004,1995SSRv...71..265A,2003A&A...411L...1W,Atwood:2009ez}.

GRBs are usually divided into two types
\citep[see][]{Kouveliotou:1993,Gehrels:2006}, distinguished primarily by
the duration of the prompt burst. Long-duration bursts, with a
duration of $\gtrsim$ \unit{2}{s}, are generally associated with
hypernova explosions in star-forming galaxies. Several nearby
long GRBs have been spatially and temporally coincident with
core-collapse supernovae as observed in the optical
\citep{Campana:2006,Galama:1998,Hjorth:2003,Malesani:2004}.
Follow-up observations by X-ray, optical, and radio telescopes of the
sky near GRBs have yielded detailed measurements of afterglows
from more than \checked{500} GRBs to date; some of these observations
resulted in strong host galaxy candidates, which allowed redshift determination
for more than \checked{200} bursts \citep{GreinerGRB}.

Short GRBs, with a duration $\lesssim$ \unit{2}{s}, are thought to
originate primarily in the coalescence of a neutron star (NS) with another
compact object \citep[see, e.g.,][and references therein]{NakarReview:2007},
such as a neutron star or black hole (BH).
There is growing evidence that finer distinctions may be drawn between
bursts \citep{Zhang:2007,Bloom:2008}; for example, it is estimated that
up to $\sim$15\% of short GRBs could be associated with soft gamma
repeaters \citep{Nakar:2006,2009MNRAS.395.1515C}, which emit bursts of X-rays and
gamma rays at irregular intervals with lower fluence than compact binary
coalescence engines \citep{Hurley:2005,Palmer:2005}.

In the compact binary coalescence model of short GRBs, a neutron star and
compact companion in otherwise stable orbit lose energy to gravitational
waves and inspiral. The neutron star(s) tidally disrupt shortly
before coalescence, providing matter, some of which is ejected in
relativistic jets.
The prompt $\gamma$-ray emission is widely thought to be created by internal
shocks, the interaction of outgoing matter shells at different
velocities, while the afterglow is thought to be created by
external shocks, the interaction of the outflowing matter with the
interstellar medium \citep{meszaros:2006,NakarReview:2007}.
If the speed of gravitational radiation equals the speed of light as
we expect, then for an observer in the cone of the collimated outflow, the
gravitational-wave inspiral signal will arrive a few seconds
before the electromagnetic signal from internal shocks.
Several semi-analytical calculations of the final stages of a NS--BH
inspiral show that the majority of matter plunges
onto the BH within \unit{1}{s} \citep{Davies:2005}.  Numerical simulations on
the mass transfer suggest a timescales of milliseconds \citep{Shibata:2007}
or some seconds at maximum \citep{Faber:2006tx}.
Also, it has been found in simulations that the vast majority of the NS
matter is accreted onto the BH directly and promptly (within hundreds of ms)
without a torus that gets
accreted later \citep{Rosswog:2006ue,Etienne:2007}.

Compact binary coalescence is anticipated to generate
strong gravitational waves in the sensitive frequency band of Earth-based
gravitational-wave detectors \citep{thorne.k:1987}.
The direct detection of gravitational waves associated with a short GRB would
provide direct evidence that the progenitor is indeed a compact
binary; with such a detection it would be possible to measure component
masses \citep{Cutler:1994,FinnChernoff:1993}, measure component
spins \citep{Poisson:1995ef}, constrain NS equations of
state \citep{flanagan:021502,Read:2009},
test general relativity in the strong-field regime \citep{Will:2005va}, and
measure calibration-free luminosity distance \citep{Nissanke:2009kt},
which is a measurement of the Hubble expansion and dark energy.

In this paper, we report on a search for gravitational-wave inspiral signals
associated with the short GRBs that occurred during the fifth science
run (S5) of LIGO \citep{Abbott:2007kv}, from 4 November 2005 to
30 September 2007, and the first
science run (VSR1) of Virgo \citep{Acernese:2008},
from 18 May 2007 to 30 September 2007.
S5 represents the combined operation of the three LIGO detectors,
one Michelson interferometer with \unit{4}{km} long orthogonal arms at Livingston, LA, USA, named L1,
and two interferometers located at Hanford, WA, USA, named H1 and H2,
with lengths of \unit{4}{km} and \unit{2}{km} respectively. VSR1 represents
the operation of the Virgo interferometer located at Cascina, Italy,
named V1, which has a length of \unit{3}{km}.
During the S5/VSR1 joint run, 212 GRBs were discovered by different
satellite missions
(39 of them during VSR1 times), 33 of which we classified as
search targets (8 of them in VSR1 times).
See section~\ref{sec:data_selection} for more details on the selected GRBs.

A similar search in the same
LIGO/Virgo data-set was performed in \citet{S5exttrigBurst},
looking for short-duration gravitational-wave bursts in association
with 137 GRBs recorded during S5/VSR1, both long and short. The analysis reported
upper limits on the strain of a generic burst of circularly polarized
gravitational
radiation, predominantly at the detectors' most sensitive frequencies.
These were translated into lower
limits in distance by assuming that $\mathunit{0.01}{M_\odot}$
is converted into isotropically emitted gravitational waves.
In contrast, the search described in this paper does not make any assumption
on the polarization of the gravitational waves and searches for
the specific signals expected from binary coalescenses. Importantly,
the present search can distinguish a coalescence signal from other
models and estimate the progenitor parameters.

The remainder of the paper is organized as follows.
In Section~\ref{sec:search}, we discuss the set of GRBs we chose
for this analysis and outline our analysis methods.
In Section~\ref{sec:results}, we present the results and
astrophysical implications for the GRBs in our sample.

\section{Search Methods}
\label{sec:search}

\subsection{Experimental setup}
\label{sec:segmentation}

\defcitealias{GehrelsPalmerPrivate}{Gehrels et al. 2008}

The binary coalescence model suggests that the time delay between the
arrival of a gravitational wave and the arrival of the subsequent
electromagnetic burst, referred to as trigger time, is a few seconds.
We assessed uncertainties in reported trigger times and
quantization in our own analysis along integer second boundaries,
finding that these each contribute less than one second.
For example, when the Swift BAT instrument determines that the count
rate has risen above a threshold, it waits for the maximum to pass,
checking with a \unit{320}{ms} cadence \citepalias{GehrelsPalmerPrivate};
it reports the start time of the block containing the maximum, rather than
making any attempt to identify the start of the burst, and does so with
a \unit{320}{ms} granularity. As another example, there have been reports
of sub-threshold precursors to many GRBs \citep{2009A&A...505..569B}.
For each GRB in our sample, we checked tens of seconds of light-curve by
eye to look for both excessive difference between the trigger time and
the apparent rise time, and also for precursors, but found nothing to
suggest that we should correct the published trigger times. The largest
timing uncertainty we identified is the delay between the compact merger
and the prompt emission of the internal shocks. We search for
gravitational-wave signals within an on-source
segment of \timewindow{} around each trigger time for each GRB of
interest, feeling that this window captures the physical model with
some tolerance for its uncertainties.

Because we believe that a gravitational wave associated with a GRB only
occurs in the on-source segment, we use \numberoff{} off-source trials,
each \seglen{} long, to estimate the distribution of
background due to the accidental coincidences of noise triggers.
We also re-analyze the off-source trials with simulated signals
added to the data to test the response of our search to signals; these
we call injection trials.
The actual number of off-source trials included in the analysis varied by
GRB, as the trials that overlapped with data-quality vetoes were
discarded \citep{abbott-2009}. To prevent biasing our background
estimation due to a potential loud signal in the on-source trial,
the off-source segments do not use data within \unit{48}{s} of the
on-source segment, reflecting the longest duration of templates in our bank.
Finally, we discard \unit{72}{s} of data subject to filter transients on
both ends of the off-source region. Taking all of these requirements into
account, the minimum analyzable time is \unit{\checked{2190}}{s}.

\subsection{Sample selection}
\label{sec:data_selection}

\begin{table}
\newcolumntype{d}{D{.}{.}{-1}}
\begin{center}
\begin{tabular}{lcdl}
\hline\hline
 GRB & Redshift & \multicolumn{1}{c}{Duration (s)} & References\\\hline
 051114 & \ldots          &  2.2 & G4272, G4275 \\
 051210 & \ldots          &  1.2 & G4315, G4321 \\
 051211 & \ldots          &  4.8 & G4324, G4359 \\
 060121 & \ldots          &  2.0 & G4550 \\
 060313 & $<1.7$ &  0.7 & G4867, G4873, G4877 \\
 060427B & \ldots          &  2.0 & G5030 \\
 060429 & \ldots          &  0.25 & G5039 \\
 061006 & \ldots          &  0.50 & G5699, G5704 \\
 061201 & \ldots          &  0.80 & G5881, G5882 \\
 061217 & 0.827 &  0.30 & G5926, G5930, G5965 \\
 070201 & \ldots          &  0.15 & G6088, G6103 \\
 070209 & \ldots          &  0.10 & G6086 \\
 070429B & \ldots          &  0.50 & G6358, G6365 \\
 070512 & \ldots          &  2.0 & G6408 \\
 070707 & \ldots          &  1.1 & G6605, G6607 \\
 070714 & \ldots          &  2.0 & G6622 \\
 070714B & 0.92 &  64.0 & G6620, G6623, G6836 \\
 070724 & 0.46 & 0.40 & G6654, G6656, G6665 \\
 070729 & \ldots          &  0.90 & G6678, G6681 \\
 070809 & \ldots          & 1.3 & G6728, G6732 \\
 070810B & \ldots          &  0.08 & G6742, G6753 \\
 070923 & \ldots          &  0.05 & G6818, G6821 \\ \hline
\end{tabular}
\end{center}
\caption{Parameters of the \numGRBs{} GRBs selected for this search.
The values in the References column give the number of the GRB Coordinates
Network (GCN) notice from which we took the preceding information
\citep{BarthelmyGCN}.}
\label{tab:listGRB}
\end{table}

X-ray and $\gamma$-ray instruments identified a total of 212 GRBs
during the S5 run, 211 with measured durations; 30 of them
have a $T_{90}$ duration smaller than \unit{2}{seconds}. $T_{90}$
is the time interval over which $90\%$ of all counts from a GRB are recorded.
While the $T_{90}$ classifies a burst as long or short, it is not a definitive
descriminator of progenitor systems.
In addition to the short GRBs, GRB~051211 \citep{gcn4359}
and GRB~070714B \citep{gcn6623} are formally long GRBs, but they have spectral
features hinting at an underlying coalescence progenitor.
GRB~061210 is another long-duration burst, but it exhibits the typical
short spikes of a short GRB \citep{gcn5904}. This gives a list of 33
interesting GRBs with which to search for an association with
gravitational waves from compact binary coalescence.

Around the trigger time of each
interesting GRB, we required \unit{\checked{2190}}{s} of multiply-coincident
data.
The detectors operated with individual duty cycles of 67\% to 81\% over
the span of the S5 and VSR1 runs.
Where more than two detectors had sufficient data, we
selected the most sensitive pair based on the average inspiral range,
because including a third, less sensitive detector does not enhance the
sensitivity greatly. The one exception was GRB~070923, described below.
In descending order of sensitivity, the detectors are H1, L1, H2, and V1.
This procedure yielded 11 GRBs searched for in H1--L1 coincident data,
9 GRBs in H1--H2, and 1 in H2--L1.

In addition to these 21, we analyze GRB~070923 because of its
sky location relative to the detectors' antenna patterns.
The antenna pattern changes with the location of a source relative to a
detector and can be expressed by the response
$\sqrt{F_+^2+F_\times^2}$, in which $F_+$ and $F_\times$ denote the
antenna-pattern functions \citep{Allen:2005fk}. A value of 1 corresponds
to an optimal
location of the putative gravitational-wave source relative to the
observatory, while a value of 0 corresponds to a source location that
will not induce any strain in the detector.
For this particular GRB, the optimal antenna response for Virgo is
around 0.7, while those for the two LIGO sites are about half that
(see Table~\ref{tab:listResults}), yielding a
comparable sensitivity in the direction of GRB~070923 for all three
of them. Data from H1, L1, and V1 were analyzed, making this the only
GRB involving triple coincidences.

Table~\ref{tab:listGRB} lists all \numGRBs{} target GRBs after applying
the selection criteria described in this section. Plausible redshifts
have been published for only three of these GRBs, placing them well outside
of our detectors' range, but short GRB redshift determinations are in general
sufficiently tentative to warrant searching for all of these GRBs.

GRB~070201 is also worth special mention. It was already analyzed
in a high-priority search because of the striking spatial coincidence of this
GRB with M31, a galaxy only $\sim$\unit{780}{kpc} from Earth. No gravitational-wave
signal was found and a coalescence scenario could be ruled out with $>$99\%
confidence at that distance \citep{GRB070201}, lending additional support
for a soft gamma repeater hypothesis \citep{2008ApJ...681.1464O}.
However, because of improvements in the analysis pipeline, we reanalyzed
this GRB and report the results in this paper. See Section~\ref{sec:070201}
for details.

\subsection{Candidate generation}
\label{subsec:candidategeneration}

We generated candidates using the standard, untriggered compact binary
coalescence search pipeline described in detail in \citet{Collaboration:2009tt}.
The core of the inspiral search involves correlating the measured data
against the theoretical waveforms expected from compact binary
coalescence, a technique called matched filtering \citep{helmstrom-1968}.
The gravitational waves from the inspiral phase, when the binary orbit
decays under gravitational-wave emission prior to merger, are
accurately modeled by post-Newtonian approximants in the band of the
detector's sensitivity for a wide range of
binary masses \citep{Blanchet:2006av}. The expected gravitational-wave
signal, as measured by LIGO and Virgo, depends on the masses ($m_\mathrm{NS},
m_\mathrm{comp}$) and spins ($\vec{s}_\mathrm{NS}, \vec{s}_\mathrm{comp}$)
of the neutron star and its companion (either NS or BH), as well as the
spatial location ($\alpha, \delta$), inclination angle $\iota$, and
polarization angle $\psi$ of the orbital axis. In general, the power of matched
filtering depends most sensitively on accurately tracking the phase
evolution of the signal. The phasing of compact binary inspiral
signals depends on the masses and spins, the time of merger, and an
overall phase.

We adopted a discrete bank of template waveforms that span a
two-dimensional parameter space (one for each component mass) such
that the maximum loss in signal to noise ratio (SNR) for a binary with
negligible spins would be $3\%$ \citep{hexabank}.
While the spin is ignored in the
template waveforms, we verify that the search can still detect binaries
with most physically reasonable spin orientations and
magnitudes with only moderate loss in sensitivity.
For simplicity, the template bank is symmetric in component
masses, spanning
the range $\mathunit{[2,\,40)}{M_\odot}$ in total mass.
The number of
template waveforms required to achieve this coverage depends on the
detector noise spectrum; for the data analyzed in this paper the number
of templates was around 7000 for each detector.

We filtered the data from each of the detectors through each template
in the bank. If the matched filter SNR exceeds a
threshold, the template masses and the time of the maximum
SNR are recorded. For a given template, threshold
crossings are clustered in time using a sliding window equal to the duration
of the template \citep{Allen:2005fk}. For each
trigger identified in this way, the coalescence phase and the
effective distance --- the distance at which an optimally oriented and
optimally located binary, with masses corresponding to those of the
template, would give the observed SNR --- are also computed. Triggers identified
in each detector are further required to be coincident with their time
and mass parameters with a trigger from at least one other
detector, taking into account the correlations between those
parameters \citep{Robinson:2008}. This significantly reduces the
number of background
triggers that arise from matched filtering in each detector
independently.

The SNR threshold for the matched filtering step was chosen differently
depending on which detectors' data are available for a given GRB\@.
If data from H1 and L1 were analyzed, the threshold for each detector
was set to 4.25, reflecting their comparable sensitivity.
If data from H1 and H2 were analyzed, the threshold of the latter detector
--- the less sensitive of the two --- was set to 3.5 to gain maximum
network sensitivity, while the threshold of the more sensitive detector, H1,
was set to 5.5 since any signal seen in H2 would be twice as loud in H1,
with some uncertainty.
In the single case of analyzing only H2--L1 data (GRB~070707) the
threshold was 4.25 for L1 and 3.5 for H2, and for the single case
of analyzing data with Virgo (GRB~070923), the threshold was set to 4.25 for all
involved detectors (H1, L1, and V1).
For comparison, a uniform SNR threshold of 5.5 was used in the untriggered
S5 search \citep{S5LowMassLV}.

We applied two signal-based tests to reduce and refine our trigger sets.
First, we computed a $\chi^2$ statistic \citep{Allen:2004}
to measure how different a trigger's SNR integrand looks from that of a real
signal; triggers with large $\chi^2$ were discarded. Second, we
applied the $r^2$ veto \citep{Rodriguez:2007} which discards triggers
depending on the duration that the $\chi^2$ statistic stays above a threshold.
The SNR and $\chi^2$ from a single detector were combined into an
effective SNR \citep{LIGOS3S4all}. The effective SNRs from the
analyzed detectors were then added in quadrature to form a single
quantity $\rho_\mathrm{eff}^2$ which provided better separation between
signal candidate events and background than SNR alone. The list of coincident
triggers at this stage are then called \textit{candidate events}.

\subsection{Ranking candidates}
\label{sec:ranking}

The distribution of effective SNRs from background and
from signals can vary strongly across the template bank,
depending most strongly on the chirp mass, a combination of the
two component masses that appears in the leading term of the
signal amplitude and phase \citep{thorne.k:1987}. For this reason,
we refine our candidate ranking with a likelihood-ratio statistic,
which we compute for every candidate in the on-source, off-source,
and injection trials.
In short, we define the likelihood ratio $L$ for a candidate $c$
to be the efficiency divided by the false-alarm probability.
The efficiency here is the probability of obtaining a candidate as loud or
louder than $c$ (by effective SNR) within the same region of template space
given a signal in the data. The efficiency is a function of
the signal parameters $m_\mathrm{comp}$ and $D$ and is marginalized over
all other parameters; it is obtained by simply counting across
injection trials. The false-alarm probability here is the probability of
obtaining a candidate as loud or louder than $c$ in the same region of
chirp mass from noise alone; it is obtained by counting across
off-source trials.

At the end of the search (i.e., Table~\ref{tab:listResults} and
Figure~\ref{fig:whitney}), we report a different false-alarm
probability. It is the fraction of off-source
likelihood ratios larger than the largest on-source likelihood ratio.

There is another noteworthy difference with respect to untriggered inspiral
searches. For background estimation, untriggered
searches use coincidences found between triggers from different detectors,
to which unphysical time-shifts greater than the light-travel time between
detector sites have been applied.
Unfortunately, H1 and H2, being co-located, share a common environmental
noise that is absent from the time-shift background
measurement. Being unable to estimate the significance of H1--H2 candidates
reliably, the untriggered search examines them with significantly greater
reservation and does not consider them at all in upper limit statements
on rates. The present search
performs its background estimation with unshifted coincidences under the
assumption that any gravitational wave signal will appear only in the
on-source trial. Thus, we regain the unconditional use of H1--H2 candidates.

\section{Results}
\label{sec:results}

\subsection{Individual GRB results}
\label{sec:resultsInd}

\begin{table}
\begin{center}
\newcolumntype{d}{D{.}{.}{-1}}
\begin{tabular}{l|cccc|c|dd}
\hline\hline
& \multicolumn{4}{c|}{Antenna response} & & \multicolumn{2}{c}{Excluded distance (Mpc)} \\
GRB & H1 & H2 & L1 & V1 & F.A.P. & \multicolumn{1}{c}{NS--NS} & \multicolumn{1}{c}{NS--BH} \\ \hline
051114 &0.56 & 0.56 & \ldots & \ldots & 1 & 2.3 & 6.2 \\
051210 &0.61 & 0.61 & \ldots & \ldots & 0.10 & 3.3 & 4.3 \\
051211 &0.53 & \ldots & 0.62 & \ldots & 0.66 & 2.3 & 8.9 \\
060121 &0.11 & \ldots & 0.09 & \ldots & 0.58 & 0.4 & 1.3 \\
060313 &0.59 & 0.59 & \ldots & \ldots & 0.16 & 1.4 & 4.3 \\
060427B &0.91 & \ldots & 0.92 & \ldots & 1 & 7.0 & 12.7 \\
060429 &0.92 & 0.92 & \ldots & \ldots & 0.21 & 4.3 & 6.2 \\
061006 &0.61 & 0.61 & \ldots & \ldots & 1 & 2.3 & 8.2 \\
061201 &0.85 & 0.85 & \ldots & \ldots & 1 & 4.3 & 10.1 \\
061217 &0.77 & \ldots & 0.52 & \ldots & 0.23 & 3.2 & 11.8 \\
070201 &0.43 & 0.43 & \ldots & \ldots & 0.07 & 3.3 & 5.3 \\
070209 &0.19 & \ldots & 0.12 & \ldots & 0.76 & 2.3 & 4.2 \\
070429B &0.99 & \ldots & 0.93 & \ldots & 0.31 & 8.9 & 14.6 \\
070512 &0.38 & \ldots & 0.51 & \ldots & 0.97 & 6.1 & 8.9 \\
070707 &\ldots & 0.87 & 0.79 & \ldots & 0.87 & 4.2 & 7.1 \\
070714 &0.28 & \ldots & 0.40 & \ldots & 0.72 & 4.2 & 2.3 \\
070714B &0.25 & \ldots & 0.38 & \ldots & 0.54 & 3.2 & 5.1 \\
070724 &0.53 & \ldots & 0.70 & \ldots & 0.84 & 5.1 & 11.8 \\
070729 &0.85 & 0.85 & \ldots & \ldots & 0.40 & 7.0 & 10.8 \\
070809 &0.30 & 0.30 & \ldots & \ldots & 1 & 2.3 & 4.3 \\
070810B &0.55 & \ldots & 0.34 & \ldots & 0.50 & 2.3 & 6.1 \\
070923 &0.32 & \ldots & 0.40 & 0.69 & 0.74 & 5.1 & 7.9 \\
\hline
\end{tabular}

\end{center}
\caption{Summary of the results for the search for gravitational waves
from each GRB. The Antenna Response column contains
the response for each detector as explained in
Section~\ref{sec:data_selection}; an ellipsis (\ldots) denotes that a detector's
data were not used. F.A.P. is the false-alarm
probability of the most significant on-source candidate for a GRB as measured
against its off-source trials, as explained in Section~\ref{sec:ranking}.
On-source trials with no candidates above threshold are assigned a F.A.P. of 1.
The last two columns show the lower limits at 90\% CL on distances,
explained in Section~\ref{sec:interpret}.}
\label{tab:listResults}
\end{table}

\begin{figure}[tbp]
\begin{center}
\includegraphics[]{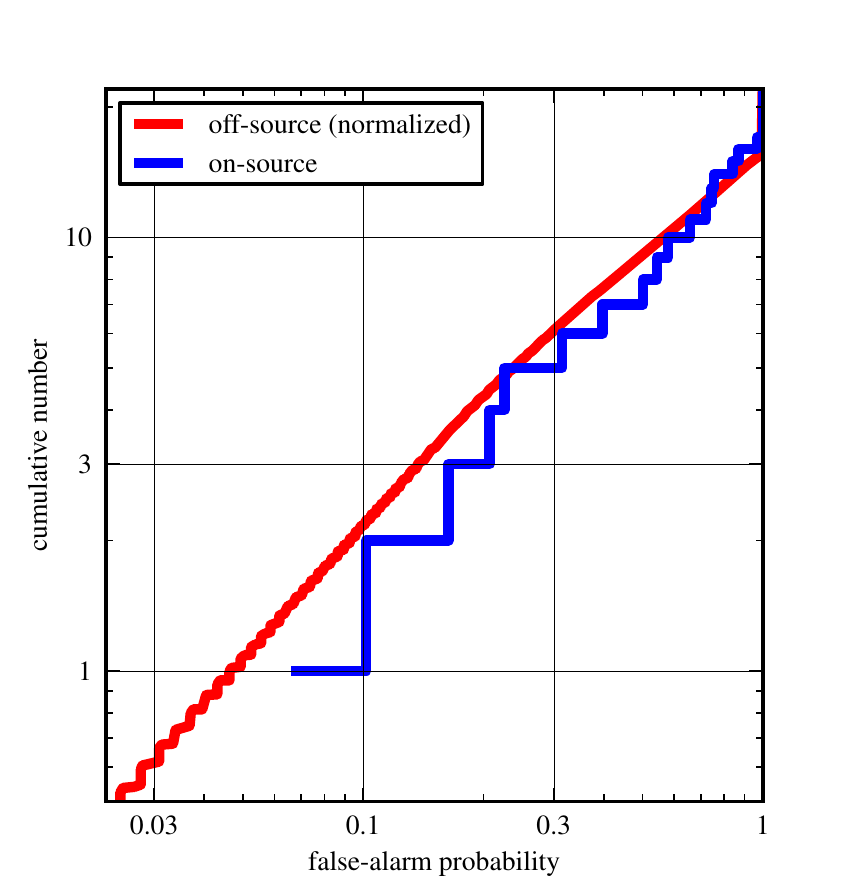}
\end{center}
\caption{Cumulative false-alarm probabilities for the
most significant candidate in each on- and off-source trial, as
described in Section~\ref{sec:ranking}.}
\label{fig:whitney}
\end{figure}

We found no evidence for a gravitational-wave signal in coincidence with any
GRB in our sample. We ran the search as described in the previous
section and found that the loudest observed candidates in each GRB's
on-source segment is consistent with the expectation from its off-source
trials. The results are summarized in Table~\ref{tab:listResults}, with
brief highlights in the following subsections. A graphical comparison of
on-source to off-source false-alarm probability is shown in
Figure~\ref{fig:whitney}.

\subsubsection{GRB~070201}
\label{sec:070201}

\renewcommand{\thefootnote}{\fnsymbol{footnote}}
The reanalysis of GRB~070201 yielded candidates in the on-source segment,
despite having no coincident candidate at all in the previous
analysis \citep{GRB070201}. This is consistent because the threshold
for H2 has been lowered from 4.0 to 3.5 and the coincident trigger found in
this reanalysis happened to lie very close to the larger threshold in
the previous search.
The reanalysis yields a false-alarm probability of 6.8\%, the smallest in
the set of analyzed GRBs
\footnote[1]{In public presentations of preliminary results, GRB~061006 was
erroneously highlighted as having the loudest candidate due to a
\unit{22.8}{s} offset in the GRB time. Swift's initial GCN alert
\citep{Schady06_GCN5699} was later corrected \citep{Schady06_GCNR6.1},
but we initially overlooked this correction.}.
This value is completely within our expectations when we consider that we
examined \numGRBs{} GRBs.

\subsubsection{GRB~070923}

GRB~070923 was the GRB for which H1, L1, and V1 had comparable
sensitivity and we accepted triggers from all three detectors.
There were no triply-coincident candidates in the on-source
trial, but there were surviving doubly-coincident candidates,
the loudest of which had a false-alarm
probability of {\statGRBLV}\%.

\subsection{Distance exclusions}
\label{sec:interpret}

\begin{figure}[tbp]
\begin{center}
\includegraphics[]{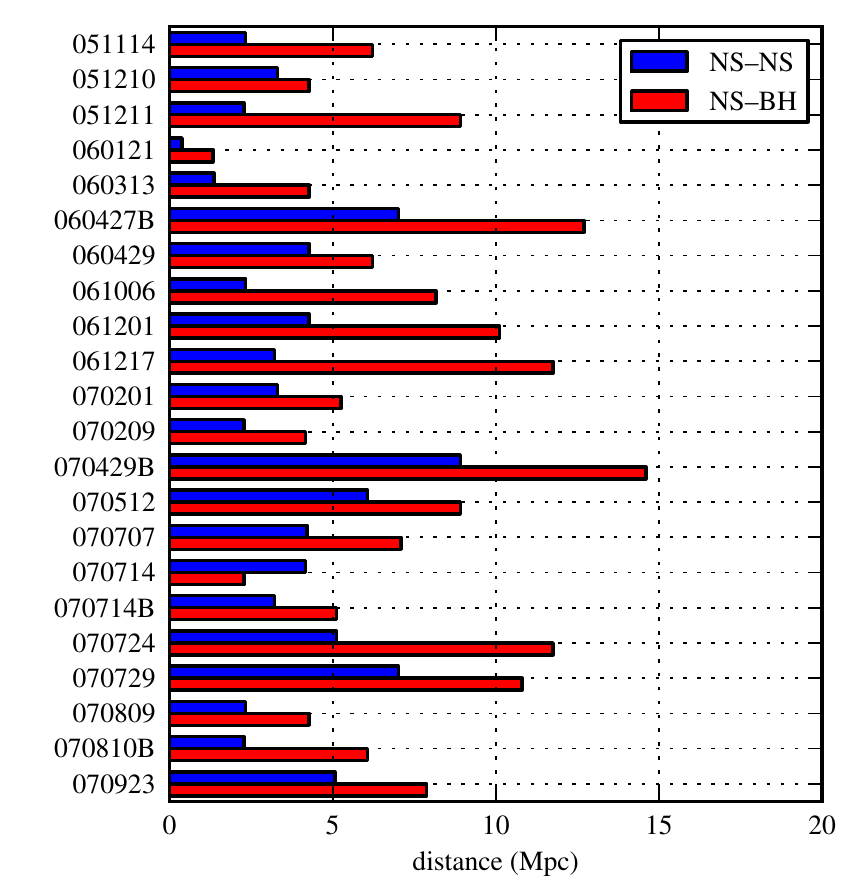}
\end{center}
\caption{Lower limits on distances at 90\% CL to putative NS--NS and
NS--BH progenitor systems, as listed in Table~\ref{tab:listResults}
and explained in Section~\ref{sec:interpret}.}
\label{fig:histRates}
\end{figure}

With our null observations and a large number of simulations,
we can constrain the distance to each GRB assuming it was caused by a
compact binary coalescence with a neutron star (with a mass in the range
$\mathunit{[1,\,3)}{M_\odot}$) and a companion of mass $m_\mathrm{comp}$.
For a given $m_\mathrm{comp}$ range, we used the approach of
\citet{Feldman:1997qc} to compute regions in distance where gravitational-wave
events would, with a given confidence, have produced results inconsistent with our
observations.
Figure~\ref{fig:histRates} shows the lower Feldman--Cousins distances for the
\numGRBs{} analyzed GRBs at 90\% confidence for two illustrative choices for
the companion mass range. The values are also listed in Table~\ref{tab:listResults}.
Because the companion mass range has been divided
into equally spaced bins, we report on a `NS--NS'
system in which the companion mass is in the range $\mathunit{[1,\,4)}{M_\odot}$
and a `NS--BH' system in which the BH has a mass in the range
$\mathunit{[7,\,10)}{M_\odot}$. The median exclusion distance for a NS--BH system is
\unit{\mediandistNSBH}{Mpc} and for a NS--NS system is
\unit{\mediandistBNS}{Mpc}.
These distances were derived assuming no beaming (uniform prior on $\cos\iota$).
NS--BH distances are typically higher than NS--NS because more massive systems
radiate more total gravitational-wave energy.
The excluded distance depends
on various parameters: the location of the GRB on the sky,
the detectors used for the GRB,
the noise floor of the data itself, and the likelihood ratio
of the loudest on-source
candidate event for the GRB\@.

We drew the simulations from a distribution in which our marginalized
parameters roughly reflect our priors on these astrophysical compact binary
systems. In our models, a signal is completely specified by $(m_\mathrm{NS},
m_\mathrm{comp}, \vec{s}_\mathrm{NS}, \vec{s}_\mathrm{comp}, \iota, \psi,
t_0, D, \alpha, \delta)$. Of these, we wish to constrain $m_\mathrm{comp}$
and $D$,
marginalizing over everything else. We drew the NS mass
$m_\mathrm{NS}$ uniformly from $\mathunit{[1,\,3)}{M_\odot}$; the magnitudes of
the NS spins $|\vec{s}_\mathrm{NS}|$ were half 0 and half uniform
in $[0,\,0.75)$ \citep{Cook:1993qr}; the magnitude of the companion's spin
$|\vec{s}_\mathrm{comp}|$ were half 0 and half uniform in $[0,\,0.98)$
\citep{Mandel:2009nx};
the orientations of the spins were uniform in solid angle; the
inclination $\iota$ of the normal to the binary's orbital plane relative
to our line of sight was conservatively chosen to be uniform in
$\cos\iota$ instead of making an assumption about the GRB beaming angle; the
polarization angle $\phi$ was uniform in $[0,\,2\pi)$;
the coalescence time $t_0$ was uniform over the off-source region;
the declination $\delta$ was set to that of the GRB; the right
ascension $\alpha$ was also set to that of the GRB, but was adjusted based
on $t_0$ to keep each simulation at the same location
relative to the detector as the GRB\@.

A number of systematic uncertainties enter into this analysis, but
amplitude calibration error and Monte--Carlo counting statistics from the
injection trials have the largest effects. We multiplied exclusion distances
by $1.28 \times (1 + \delta_\mathrm{cal})$, where $\delta_\mathrm{cal}$
is the fractional uncertainty (10\% for H1 and H2; 13\% for L1; 6\% for V1
\citep{Marion:2008}). The factor of 1.28 corresponds to a 90\%
pessimistic fluctuation, assuming Gaussianity. To take the counting statistics
into account, we stretched the Feldman--Cousins confidence belts to cover
the probability $\mathrm{CL} + 1.28 \sqrt{\textrm{CL} (1-\textrm{CL})/n}$,
where CL is the desired confidence limit and $n$ is the number of
simulations contained in the $(m_\mathrm{comp},\,D)$ bin for which we are
constructing the belt.

\subsection{Population statement}

In addition to the individual detection searches above, we would
like to assess the presence of gravitational-wave signals that are
too weak to stand out above background separately, but that are
significant when the entire population of analyzed GRBs is taken together.
We compare the cumulative distribution of the false alarm 
probabilities of the on-source sample with the off-source sample.
The on-source sample consist of the results of all \numGRBs{} individual
searches, including those for GRBs with known redshifts, and the
off-source sample consists of \numbertotaloff{}
results from the off-source trials. This number is lower than
$\numGRBs{}\times\numberoff{}$ because for some GRBs, some off-source
trials were discarded due to known data quality issues.

These two distributions are compared in Figure~\ref{fig:whitney}. To
determine if they are consistent with being drawn from the same parent
distribution, we employ the non-parametric
Wilcoxon--Mann--Whitney U statistic \citep{MannWhitney}, which is a
measure of how different two populations are.
Applying the U test, we find
that the two distributions are consistent with each other; if the
on-source and off-source significances were drawn from the same
distribution, they would yield a U statistic greater than what we
observed \mwuprob\% of the time.
Therefore we find no evidence for an excess of weak gravitational-wave
signals associated with GRBs.

\section{Discussion}

We searched data taken with the three LIGO detectors and the Virgo
detector for gravitational-wave signatures of compact binary coalescences
associated with \numGRBs{} GRBs but found none. We were sensitive to
systems with total masses $\mathunit{2}{M_\odot} < m
< \mathunit{40}{M_\odot}$. We also searched for a population of signals too
weak to be individually detected, but again found no evidence.
While there are few redshift determinations for short GRBs, it appears that
the distribution is peaked around $\langle z \rangle \sim 0.25$ \citep{NakarReview:2007},
far outside initial detector sensitivity,
so it is not surprising that the S5/VSR1 run yielded no
detections associated with short GRBs.

\acknowledgments
We are indebted to the observers of the electromagnetic events and the GCN for providing us with valuable data.
The authors gratefully acknowledge the support of the United States
National Science Foundation for the construction and operation of the
LIGO Laboratory, the Science and Technology Facilities Council of the
United Kingdom, the Max-Planck-Society, and the State of
Niedersachsen/Germany for support of the construction and operation of
the GEO600 detector, and the Italian Istituto Nazionale di Fisica
Nucleare and the French Centre National de la Recherche Scientifique
for the construction and operation of the Virgo detector. The authors
also gratefully acknowledge the support of the research by these
agencies and by the Australian Research Council, the Council of
Scientific and Industrial Research of India, the Istituto Nazionale di
Fisica Nucleare of Italy, the Spanish Ministerio de Educaci\'on y
Ciencia, the Conselleria d'Economia Hisenda i Innovaci\'o of the
Govern de les Illes Balears, the Foundation for Fundamental Research
on Matter supported by the Netherlands Organisation for Scientific
Research, the Royal Society, the Scottish Funding Council, the
Scottish Universities Physics Alliance, The National Aeronautics and
Space Administration, the Carnegie Trust, the Leverhulme Trust, the
David and Lucile Packard Foundation, the Research Corporation, and
the Alfred P. Sloan Foundation.

This document bears the LIGO document number P0900074.

\bibliographystyle{apj}
\bibliography{../bibtex/iulpapers}

\end{document}